\pdfoutput=1

\documentclass{article}

\usepackage[utf8]{inputenc}     
\usepackage[margin=10pt,font=small,labelfont=bf,labelsep=endash]{caption}
\usepackage[T1]{fontenc}
\usepackage{setspace}
\usepackage{booktabs}
\usepackage{multirow}
\usepackage{tabularx}
\usepackage{paralist}
\usepackage{ifthen}
\usepackage{listings}
\usepackage{amsmath}
\usepackage{amsfonts}
\usepackage{amssymb}
\usepackage{color}
\usepackage[table]{xcolor}
\usepackage{algorithm}
\usepackage{algpseudocode}
\usepackage{url}
\usepackage{rotating}
\usepackage{subfig}
\usepackage{arydshln}
\usepackage{todonotes}
\usepackage[normalem]{ulem}
\usepackage{siunitx}
\usepackage{diagbox}
\usepackage[htt]{hyphenat}
\usepackage{enumitem}
\usepackage{tikz}
\usepackage{pgfplots}
\usetikzlibrary{plotmarks}
\usepackage{authblk}
\usepackage{float} 
\usepackage{pdfpages}
\usepackage{fancyvrb}
\usepackage[most]{tcolorbox}
\usepackage[top=3cm, bottom=3cm, inner=3.5cm, outer=2.5cm]{geometry}
\usepackage[absolute]{textpos}

\graphicspath{{images/}}

\newcommand{\progvar}{\texttt}
\newcommand{\progfunc}{\textnhtt}
\newcommand\ooclass[1]{{\normalfont\fontfamily{cmvtt}\selectfont #1}}
\newcommand\oointerface[1]{{\itshape\fontfamily{cmvtt}\selectfont #1}}

\title{cf4ocl: a C framework for OpenCL}

\author[1]{Nuno Fachada}
\author[2]{Vitor V. Lopes}
\author[3]{Rui C. Martins}
\author[1]{Agostinho C. Rosa}

\affil[1]{Institute for Systems and Robotics (ISR/IST), LARSyS, Instituto Superior Técnico, Av. Rovisco Pais, 1, 1049-001 Lisboa, Portugal}
\affil[2]{UTEC - Universidad de Ingeniería \& Tecnología, Lima, Jr. Medrano Silva 165, Barranco, Lima, Perú}
\affil[3]{INESC TEC, Campus da FEUP, Rua Dr. Roberto Frias, 4200-465 Porto, Portugal}

\providecommand{\keywords}[1]{\textbf{\textit{Keywords---}} #1}

\date{}

\begin{document}

\begin{textblock*}{210mm}(3mm,3mm)
	\noindent The peer-reviewed version of this paper is published in Science of Computer Programming (\url{https://doi.org/10.1016/j.scico.2017.03.005}). This version is typeset by the authors and differs only in pagination and typographical detail.
\end{textblock*}

\maketitle

\begin{abstract}

OpenCL is an open standard for parallel programming of heterogeneous compute devices, such as GPUs, CPUs, DSPs or FPGAs. However, the verbosity of its C host API can hinder application development. In this paper we present cf4ocl, a software library for rapid development of OpenCL programs in pure C. It aims to reduce the verbosity of the OpenCL API, offering straightforward memory management, integrated profiling of events (e.g., kernel execution and data transfers), simple but extensible device selection mechanism and user-friendly error management. We compare two versions of a conceptual application example, one based on cf4ocl, the other developed directly with the OpenCL host API. Results show that the former is simpler to implement and offers more features, at the cost of an effectively negligible computational overhead. Additionally, the tools provided with cf4ocl allowed for a quick analysis on how to optimize the application.

\end{abstract}



\section{Introduction}
\label{sec:intro}

OpenCL is an open standard for parallel programming of heterogeneous compute devices, such as graphics processing units (GPUs), central processing units (CPUs), digital signal processors (DSPs) or field-programmable gate arrays (FPGAs). OpenCL is divided in two parts: a) a C99-based language\footnote{A C++14-based kernel language is introduced in OpenCL 2.1 \cite{ocl21released2015}.} for programming these devices; and, b) a C application programming interface (API) to launch device programs and manage device memory from a host processor \cite{opencl2014}.

One of the problems often associated with OpenCL is the verbosity of its C host API \cite{oclprog2014,smalloclquest2014,said2016hicl}. Even the simplest of programs requires a considerable amount of repetitive, error-prone boilerplate code. However, the upside is that the API is very flexible and offers fine-grained control of all aspects concerning the development of parallel programs.

In this paper we present the C Framework for OpenCL, cf4ocl, a software library with the following goals: 1) promote the rapid development of OpenCL host programs in C (with support for C++), while avoiding the associated API verbosity; 2) assist in the benchmarking of OpenCL events, such as kernel execution and data transfers; and, 3) simplify the analysis of the OpenCL environment and of kernel requirements. Summarizing, cf4ocl allows the programmer to focus on device code, which is usually more complex and what delivers the end results.

This article is organized as follows. In Section \ref{sec:background}, we describe the problem cf4ocl is trying to solve and discuss alternative libraries with similar goals. In Section \ref{sec:softframe}, the architecture and functionality of the library are presented. Implementation details are discussed in Section \ref{sec:impl}. An example application for massive pseudo-random number generation is introduced in Section \ref{sec:examples}. Two realizations of this application, implemented in pure OpenCL and using cf4ocl, respectively, are compared and analyzed in Section \ref{sec:results}. This paper closes with Section \ref{sec:conclusions}, where we provide a number of conclusions regarding the advantages provided by cf4ocl.

\section{Problems and Background}
\label{sec:background}

The OpenCL host API presents a low-level abstraction of the underlying computational platforms. It is well designed, well organized, offering maximum flexibility while supporting a large variety of compute devices. However, this flexibility comes at a cost. As a low-level C API, each function performs a very specific task. Built-in functionality is scarce: there is no error handling or automatic resource management. The end-user must implement this functionality himself. Thus, the simplest of OpenCL host programs requires a considerable amount of repetitive and error-prone boilerplate code. 

The problem is minimized when using bindings or wrappers for other programming languages, which abstract away much of the associated complexity. Examples include the official C++ bindings \cite{oclcppbind2016}, and third-party bindings for languages such as Python \cite{klockner2012pycuda}, Haskell \cite{gaster2013embedding}, Java \cite{ocljava2010} or R \cite{rocl2015}. In the case of C++, and in spite of the official bindings, the number of wrappers and abstraction libraries is remarkable \cite{clu2016,lawlor2011embedding,vexcl2016,szuppe2016boost,Yalamanchili2015,easycl2016,steuwer2011skelcl,vinas2015improving,goopax2014}. These libraries aim for a number of goals, such as rapid and/or simplified development of OpenCL programs, high-level abstractions for common computation and communication patterns, embedded OpenCL kernel code within C++ programs or handling of multiple OpenCL platforms and devices.

In turn, there are a number of libraries for developing OpenCL programs in pure C host code. For instance, Simple OpenCL \cite{simpleopencl2013} aims to reduce the host code needed to run OpenCL C kernels. It offers a very high-level abstraction, with a minimum of two types and three functions required to execute OpenCL code on a single compute device. While simple, it can be inflexible in more complex workflows, e.g., involving multiple command queues or devices.

The OpenCL utility library \cite{oclutil2012} provides a set of functions and macros to make the host side of OpenCL programming less tedious. The functions and macros perform common complex or repetitive tasks such as platform and device selection, information gathering, command queue management and memory management. The library works at a lower level than Simple OpenCL, directly exposing OpenCL types. However, it is oriented towards using Pthreads \cite{lewis1998multithreaded}, privileging a single device, context and command-queue per thread. Additionally, the library has a not insignificant performance cost, as it performs a number of queries upfront.

OCL-MLA \cite{oclmla2013} is a set of mid-level abstractions to facilitate OpenCL development. It covers more of the OpenCL host API than both Simple OpenCL and OpenCL utility library, and sits in between the two concerning the level of abstraction. However, like Simple OpenCL, it is also oriented towards basic workflows focusing on a single device, context and command-queue. It features compile-time logical device configuration, management of several OpenCL object types, profiling capabilities, helper functions for manipulation of events, and a number of utilities for program manipulation. Furthermore, it offers Fortran bindings. 

Oclkit is a small wrapper library focused on platform, context, command queue and program initialization, avoiding the boilerplate code associated with these tasks \cite{oclkit2016}. It is a low-level thin wrapper, allowing the programmer to keep full control over the OpenCL workflow. Nonetheless, it does not provide much functionality beyond these tasks and associates each device with a command queue, limiting its applicability.

Finally, hiCL consists of a C/C++ and a Fortran 90 wrapper which prioritizes memory management between different computation devices, namely those sharing RAM, such as CPUs with integrated GPUs \cite{said2016hicl}. It provides a high-level abstraction for launching kernels while internally managing dependencies between kernels, devices and memory objects. However, it also associates a device with a command queue.

All of these libraries present some type of limitation, and many have seen development stall in the last few years. For example, at the time of writing, none of them supports the new approach for command queue creation introduced in OpenCL 2.0 \cite{opencl2014}.

\section{Software Framework }
\label{sec:softframe}

\subsection{Software Architecture}
\label{sec:softframe:softarch}

Cf4ocl is divided in two major components, namely the \emph{library} and the \emph{utilities}. The \emph{library} component is organized into several modules. The majority of modules are wrappers for the OpenCL API, simplifying its use. Other modules provide mechanisms for querying and selecting devices, converting OpenCL error codes into human-readable strings, managing the OpenCL platforms available in a system and profiling OpenCL programs.

The \emph{utilities} component is composed of three standalone command-line applications, which complement the functionality provided by the library. The \texttt{ccl\_devinfo} utility can be used to query OpenCL platforms and devices. The \texttt{ccl\_c} application performs offline compilation, linking and analysis of OpenCL kernels. Finally, the \texttt{ccl\_plot\_events} script plots a queue utilization chart of OpenCL commands using profiling information generated with the library component.

\subsection{Software Functionalities}
\label{sec:softframe:softfunc}

Cf4ocl presents the developer with an object-oriented interface which wraps and extends the functionality provided by the OpenCL C API, offering a number of features:

\begin{itemize}
	\item Simplified memory management:
	\begin{itemize}
		\item Clear set of constructor and destructor functions for all objects.
		\item Automatic memory management for intermediate objects, such as information tokens retrieved from the underlying OpenCL objects.
	\end{itemize}
	\item Flexible device selection mechanism, with direct functions for common use cases and an accessible API for more complex workflows.
	\item Straightforward event dependency system, with automatic memory management of all event objects.
	\item Comprehensive error reporting.
	\item Abstracts differences in the OpenCL version of the underlying platforms, presenting a consistent API to the developer.
	\item Integrated profiling, with basic and advanced functionality.
	\item Versatile device query utility, capable of customized queries.
	\item Offline kernel compiler, linker and analyzer.
\end{itemize}

\section{Implementation}
\label{sec:impl}

\subsection{Common patterns}
\label{sec:impl:commonpatt}

As already described, cf4ocl, presents the developer with an object-oriented interface which wraps and extends the functionality provided by the OpenCL C API. While C does not directly provide object-oriented constructs, it is possible to implement features such as inheritance, polymorphism or encapsulation \cite{schreiner1993object}. Using this approach, cf4ocl is able to offer a clean and logical class system for OpenCL development in the context of a simple, compact and close-to-hardware programming language.

Each cf4ocl class is defined by a source (.c) file and a header (.h) file. The former contains the private class properties and the method implementations, while the latter defines its public API.
Methods are implemented as functions which accept the object on which they operate as the first parameter.

The constructor and destructor functions for managing cf4ocl objects are consistently named, containing the class name followed by \texttt{new} or \texttt{destroy}. Object lifecycle management is simple: for each invoked constructor, the respective destructor must also be invoked. This might seem obvious, but in several cases objects are obtained using non-constructor methods during the course of a program. These objects are automatically released and should not be destroyed by client code.

Error-throwing cf4ocl functions report errors using two approaches: a) via the return value; and, b) by initializing an optional object passed as the last argument to the function. Developers can use the method most appropriate for the program being developed, and ignore the unused approach. The first method is more limited since it only signals that an error has occurred, not providing any additional information, while the second method is more flexible. 

\subsection{The wrapper modules}

\begin{figure}[t]
\centering
\includegraphics[width=1\linewidth]{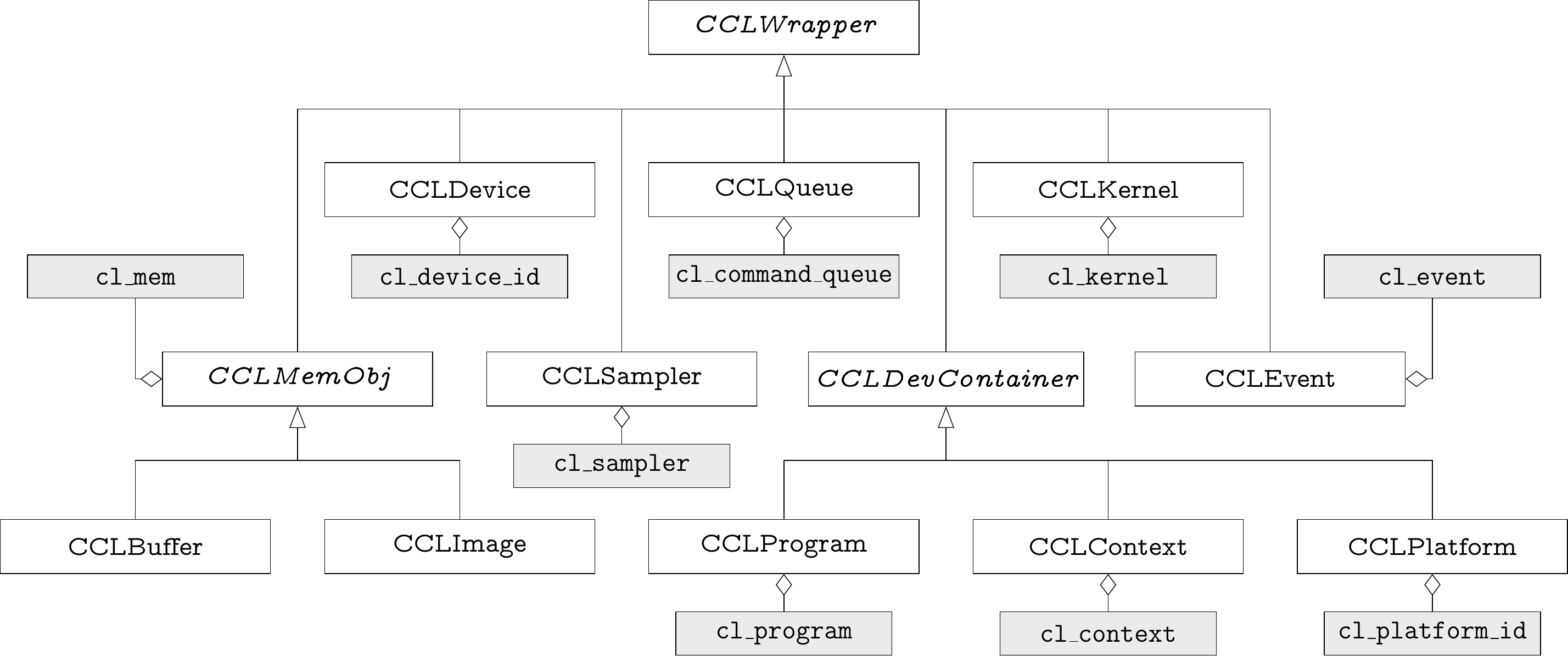}
\caption{UML diagram showing the cf4ocl class hierarchy and the one-to-one aggregation relation of each wrapper with the corresponding OpenCL object. Cf4ocl and OpenCL classes are shown in white and light gray backgrounds, respectively. Italic names correspond to abstract classes.}
\label{fig:ccluml}
\end{figure}

The several wrapper modules and their similarly named classes have an approximate one-to-one correspondence with raw OpenCL objects. These classes automatically manage the majority of intermediate memory objects, such as events and information queries, reducing lines of code (LOCs) and promoting faster, less error-prone development of client applications. Since cf4ocl follows the workflow logic of the OpenCL host API, it is straightforward for a developer to move between the two systems. Additionally, because raw OpenCL objects are always accessible to developers, a mix of OpenCL host code and cf4ocl code is possible. Developers can completely avoid any direct OpenCL calls\footnote{With the exception, at the time of writing, of few functions not yet wrapped, mostly introduced in OpenCL 2.0 and 2.1.} by using cf4ocl to its full capabilities, or use only the cf4ocl functionality that suits them. The cf4ocl class hierarchy and the relation of each class with the corresponding OpenCL object is shown in Figure \ref{fig:ccluml}.

The \oointerface{CCLWrapper} abstract super class is responsible for common functionality of wrapper objects, namely: a) wrapping/unwrapping of OpenCL objects and maintaining a one-to-one relationship between wrapped and wrapper objects; b) low-level memory management (allocation and deallocation); and, c) information handling, i.e., managing data returned by the several \progfunc{clGet*Info()} OpenCL functions.

The intermediate \oointerface{CCLDevContainer} class provides functionality for managing a set of \ooclass{CCLDevice} wrapper instances, abstracting code common to the \ooclass{CCLPlatform}, \ooclass{CCLContext} and \ooclass{CCLProgram} classes, all of which internally keep a set of devices. 

The relationship between the \oointerface{CCLMemObj} abstract class and the \ooclass{CCLBuffer} and \ooclass{CCLImage} classes follows that of the respective OpenCL types. In other words, both OpenCL images and buffers are memory objects with common functionality, and cf4ocl directly maps this relationship with the respective wrappers.

\subsection{The profiler module}
\label{sec:impl:prof}

This module provides integrated profiling of OpenCL events, such as kernel execution and data transfers. The use of this functionality with cf4ocl only requires minor additions to the application logic, 
which is not the case when directly using the OpenCL host API. It is necessary to create a \ooclass{CCLProf} object, and after all the computations and memory transfers have taken place, pass it the utilized \ooclass{CCLQueue} wrappers, and order it to perform the profiling analysis. Naturally, this requires that command queues are created with the appropriate OpenCL profiling flag.

At this stage, different types of profiling information become available, and can be directly accessed or iterated over:

\begin{description}
	\item[Aggregate event information] Absolute and relative durations of all events with same name, represented by the \ooclass{CCLProfAgg} class. If an event name is not set during the course of the computation, the aggregation is performed by event type, i.e., by events which represent the same command.

	\item[Non-aggregate event information] Event-specific information, represented by the \ooclass{CCLProfInfo} class, such as event name (or type, if no name is given), the queue the event is associated with, and the several event instants.

	\item[Event instants] Specific start and end event timestamps, represented by the \ooclass{CCLProfInst} class.

	\item[Event overlaps] Information about event overlaps, represented by the \ooclass{CCLProfOverlap} class. Event overlaps can only occur when more than one queue is used with the same device.
	
\end{description}

While this information can be subject to different types of examination by client code, the profiler module also offers functionality which allows a more immediate interpretation of results, namely it can: a) generate a text summary of the profiling analysis; and, b) export a table of \ooclass{CCLProfInfo} data, containing queue name, start instant, end instant and event name, to a text file which can be opened with the \verb|ccl_plot_events| utility to plot a queue utilization chart of the performed computation.

\subsection{Other modules}

The remaining modules complement the functionality provided by cf4ocl. These modules are not commonly used by client code, but can be useful in certain circumstances.

The device selector module offers a filtering mechanism for selecting OpenCL devices. It is mainly used by the context wrapper module for the purpose of context creation. Nonetheless, this device filtering functionality can be used in cases for which it makes sense to select or enumerate devices depending on their characteristics, such as type (CPU, GPU or other accelerators) or vendor. The mechanism can be extended by client code via plug-in filters.

The device query module provides much of the functionality of the \texttt{ccl\_devinfo} utility in library form. While its main goal is to support this utility, it may also be of use to client code.

The errors module contains a single function which converts OpenCL error codes into human-readable strings. It is used by all error-throwing cf4ocl functions, but can also be useful in situations where client code simply requires conversion of error codes to strings.

The platforms module offers functionality for managing the OpenCL platforms available in the system. It is different from the platform wrapper module, since it works with the set of available platforms, and not with the platform objects themselves.

\section{Example: massive pseudo-random number generator}
\label{sec:examples}

Pseudo-random number generators (PRNGs) are commonly used as a source of randomness and/or noise generation in scenarios such as simulations, Monte Carlo methods, computer graphics, genetic and evolutionary algorithms or artificial neural networks \cite{coddington1997random,neves2012fast}. It is often required or preferable to keep the computational cost of randomization as low as possible \cite{langdon2008fast,thomas2009comparison,zhmurov2010generation}. Thus, a plausible strategy for random number consuming CPU-bound applications would be to use the GPU as a coprocessor for generating the required random number stream(s). One of the methods available in the GASPRNG CUDA library follows this approach \cite{gao2013gasprng}. In order to illustrate the functionality of cf4ocl, we provide an example implementing this technique.

Our example is a standalone program which outputs random numbers in binary format to the standard output (\textit{stdout}). The generated random number stream can be redirected to consumer applications using operating system pipes. The program accepts two parameters: a) $n$, the quantity of 64-bit (8-byte) random values to generate per iteration; and, b) $i$, the number of iterations producing random values. The total number of random bytes generated during program execution, $N$, is given by Eq. \ref{eq:totranvals}:

\begin{equation}
\label{eq:totranvals}
N=8 \cdot ni
\end{equation}

\begin{figure}[t]
\centering
\includegraphics[width=1\linewidth]{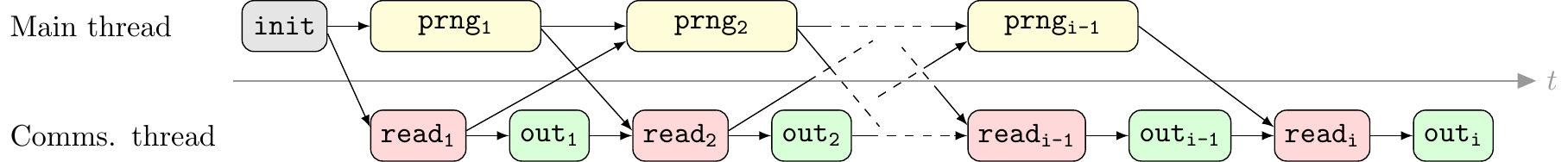}
\caption{Program threads and execution blocks. The \texttt{init} and \texttt{prng} blocks represent the initialization and PRNG kernels, respectively. In turn, \texttt{read} blocks depict device-host data transfers, while \texttt{out} blocks symbolize output of random values to \textit{stdout}. Blocks pointed to by arrows can only execute when preceding block(s) have finished execution. Block sizes do not correspond to real execution times.}
\label{fig:exampleprng}
\end{figure}

For example, the following command\footnote{Given for UNIX-style platforms, e.g., Linux, macOS or MSYS/Cygwin (Windows).} executes our program with $2^{24}$ 64-bit random numbers per iteration during \num{10000} iterations, and redirects the output to the Dieharder PRNG test suite \cite{brown2013dieharder}:

\begin{verbatim}
$ ./rng_ccl 16777216 10000 | dieharder -g 200 -a
\end{verbatim}

The program uses two threads: the main thread and the communications thread. The initialization and PRNG kernels run in the former, while the device-host data transfers and stream output to \textit{stdout} are performed in the latter, as shown in Figure~\ref{fig:exampleprng}. The threads are instantiated with the Pthreads API and are synchronized using semaphores. Each thread is associated with one OpenCL command queue.

At the device level, each work-item\footnote{i.e., a thread in OpenCL terminology.} generates a 64-bit random value per invocation. The \texttt{init} kernel creates the initial random values by applying a hash function \cite{wang1997inthash} to the global ID of the associated work-items. The generated values not only constitute the first batch of random numbers, but also serve as seeds for the next batch. When the \texttt{init} kernel terminates, the \texttt{prng} kernel can start generating the next batch of random values, while the \texttt{read} operation fetches the initial numbers produced by the \texttt{init} kernel. In subsequent iterations, the \texttt{read} operation will fetch random values generated by the previous iteration of the \texttt{prng} kernel. This is possible using a device-side double-buffering approach. While \texttt{prng} and \texttt{read} processes read from one buffer, the former writes the new batch of values to a second buffer. Buffers are then swapped, and both processes can start a new iteration without major interruptions.

The example is conceptually simple, but incorporates relatively complex patterns such as simultaneous kernel execution and data transfers using multiple command queues. This approach allows us to compare a pure OpenCL version with a cf4ocl realization. This example would be difficult to implement with the majority of other libraries, which tend to associate one device with a single command queue.

While the example is adequate for the stated purposes, it has certain limitations for real world use. First, it uses a simple Xorshift PRNG \cite{marsaglia2003xorshift} that, while fast, is probably not ideal for being partitioned in parallel contexts \cite{coddington1997random}. Second, a host-side dual buffer implementation would further improve performance by allowing concurrent \texttt{read} and \texttt{out} processes using an additional host thread. Third, the PRNG kernel does not use vectorization, which would allow individual work-items to generate more than one random value per invocation.

\section{Empirical results}
\label{sec:results}

In this section we compare two implementations of the example described in Section~\ref{sec:examples}. The first realization is directly coded with the OpenCL C host API (Listing~S1), while the second uses cf4ocl (Listing~S2). The two implementations share a compatibility header for cross-platform semaphore usage (Listing~S3), and execute the same kernel code (Listings~S4 and S5). Besides being provided as Supplementary material (Listings~S1--S5), the complete source code of the example is also part of the \texttt{cf4ocl-examples} repository available at \url{https://github.com/fakenmc/cf4ocl-examples}. The implementations are compared from two perspectives: a) code complexity; and, b) performance.

\subsection{Comparison of code complexity}

The number of LOCs offers a first impression of the work required to implement the example. By LOCs we mean physical lines of code, so this excludes blank lines and comments. 
While the pure OpenCL implementation requires 290 LOCs, the cf4ocl version is about 37\% smaller, needing 183 LOCs. A minimum-LOC approach which guarantees correct behavior is followed in the former realization. However, when qualitatively compared with the latter, it lacks more detailed profiling (e.g., overlap detection), user-friendly error messages and flexible kernel work size calculation.

Nonetheless, LOCs may not be by themselves a direct measure of programmer productivity. It is also important to understand what blocks of code are simplified, and how the introduced abstractions facilitate or encumber an understanding of the underlying OpenCL calls. Since cf4ocl follows the overall logic of the OpenCL host API, comparing the source code of the two implementations is relatively straightforward. We will perform this comparison in the following paragraphs, highlighting areas where the use of cf4ocl is particularly advantageous.

First, cf4ocl requires fewer auxiliary variables for typical OpenCL operations (e.g., device selection, object queries, loading of kernel source codes or profiling), freeing the developer of their micromanagement (i.e., creation and destruction). Analyzing the source codes it is possible to confirm that the pure OpenCL version uses considerably more of these variables (Listing~S1, lines 149--190) than the cf4ocl realization (Listing~S2, lines 141--154).

Device selection and context creation, as well as the retrieval of information about these objects, are areas where the OpenCL host API is distinctly verbose, as documented in lines 217--265 of Listing~S1. In turn, cf4ocl provides a number of helper functions and macros which simplify this process, as shown in the following code from Listing~S2:

\begin{lstlisting}[firstnumber=181, belowskip=4pt]
/* Setup OpenCL context with GPU device. */
ctx = ccl_context_new_gpu(&err);
\end{lstlisting}
\begin{lstlisting}[firstnumber=185, belowskip=4pt, aboveskip=0pt]
/* Get device. */
dev = ccl_context_get_device(ctx, 0, &err);
\end{lstlisting}
\begin{lstlisting}[firstnumber=189, aboveskip=0pt]
/* Get device name. */
dev_name = ccl_device_get_info_array(dev, CL_DEVICE_NAME, char*, &err);
\end{lstlisting}

\noindent The \progvar{err} variable represents the error handling object described in Section~\ref{sec:impl:commonpatt}. For clarity, error handling macros are not displayed in the code snippets presented in this section.

The program creation functions offered by the OpenCL host API require kernel sources or binaries to be passed as strings. There is no native functionality for loading kernel code directly from files. Furthermore, as in the case of other information queries, obtaining the build log is also a long-winded exercise. The process of loading kernel files, program creation and building, as well as of getting the build log in case of error is shown in lines 278--329 of Listing~S1. Conversely, cf4ocl automatizes this process as shown in the following lines from Listing~S2:

\begin{lstlisting}[firstnumber=199, belowskip=4pt]
/* Create program. */
prg = ccl_program_new_from_source_files(ctx, 2, kernel_filenames, &err);
\end{lstlisting}
\begin{lstlisting}[firstnumber=203, belowskip=4pt, aboveskip=0pt]
/* Build program. */
ccl_program_build(prg, NULL, &err);
\end{lstlisting}
\begin{lstlisting}[firstnumber=206, belowskip=4pt, aboveskip=0pt]
/* Print build log in case of error. */
if ((err) && (err->code == CL_BUILD_PROGRAM_FAILURE)) {
	bldlog = ccl_program_get_build_log(prg, &err_bld);
\end{lstlisting}
\begin{lstlisting}[firstnumber=210, belowskip=4pt, aboveskip=0pt]
	fprintf(stderr, "Error building program: \n%s", bldlog);
\end{lstlisting}
\begin{lstlisting}[firstnumber=212, aboveskip=0pt]
}
\end{lstlisting}

\noindent The code is straightforward and easily readable, especially when compared with the equivalent in Listing~S1.

Going forward, it is important in terms of performance to determine appropriate kernel work sizes for the selected device. There are two fundamental work sizes to be considered when executing OpenCL kernels: global and local. The global work size (GWS) is the total number of work-items executing a given kernel. Since work-items are grouped into work-groups, the local work size (LWS) is the number of work-items per work-group. Commonly, individual work-groups are associated with and processed by a single compute unit (CU) on the selected device. Work-items are, in turn, executed on the processing elements (PEs) of the CU. As such, it is essential that the LWS is well adjusted to the CUs' capabilities, namely to the PEs it contains. The OpenCL host API allows to query device and kernel objects for determining maximum and preferred LWSs, respectively\footnote{More specifically, the preferred work-group size should be a multiple of the value returned by the kernel query.}. However, the latter query is only available from OpenCL 1.1 onwards. Also, for OpenCL versions prior to 2.0, the GWS must be a multiple of the LWS. For OpenCL 2.0 and higher, this is not mandatory, and the work-group with the highest ID may have less work-items than the remaining work-groups. In our example, the pure OpenCL implementation uses the kernel object to get the preferred LWS (lines 344--354 of Listing~S1), but this only works with OpenCL 1.1 or higher, and the performed calculations are not applicable to multiple dimensions. cf4ocl's \progfunc{ccl\_\allowbreak{}kernel\_\allowbreak{}suggest\_\allowbreak{}worksizes()} function accounts for these issues and is simpler to use, as shown in the following lines of Listing~S2:

\begin{lstlisting}[firstnumber=221, belowskip=4pt]
/* Determine preferred work sizes for each kernel. */
ccl_kernel_suggest_worksizes(kinit, dev, 1, &rws, &gws1, &lws1, &err);
\end{lstlisting}
\begin{lstlisting}[firstnumber=224, aboveskip=0pt]
ccl_kernel_suggest_worksizes(krng, dev, 1, &rws, &gws2, &lws2, &err);
\end{lstlisting}

\noindent Here, \progvar{kinit} and \progvar{krng} are the initialization and PRNG kernel objects, respectively, and \progvar{rws} is the quantity of pseudo-random numbers to generate per iteration (i.e., the real work size). The \progvar{gws\textit{i}} and \progvar{lws\textit{i}} variables, with index $i$ denoting the kernel which the work sizes refer to, are populated with appropriate GWSs and LWSs, respectively.

Setting kernel arguments is another tedious aspect of OpenCL host-side programming in C, since each individual argument must be set separately using the \progfunc{clSetKernelArg()} function. For example, lines 389--399 of Listing~S1 show, for the pure OpenCL implementation, how the arguments of the \progvar{kinit} kernel are set and how the kernel is invoked. However, the cf4ocl version of the example performs these operations with a single function call:

\begin{lstlisting}[firstnumber=256]
/* Invoke kernel for initializing random numbers. */
evt_exec = ccl_kernel_set_args_and_enqueue_ndrange(kinit, cq_main, 1, NULL,
  (const size_t*) &gws1, (const size_t*) &lws1, NULL, &err,
  bufdev1, ccl_arg_priv(bufs.numrn, cl_uint), /* Kernel arguments. */
  NULL);
\end{lstlisting}

\noindent The \progvar{evt\_exec} variable represents the event generated by launching the kernel, and \progvar{cq\_main} is the command queue associated with the main thread. The parameters in line 259 are the arguments passed to the \progvar{kinit} kernel, namely the PRNG states vector (\progvar{bufdev1}) and number of states it contains (\progvar{bufs.numrn}). Note that: a) private kernel arguments are wrapped with the \progfunc{ccl\_arg\_priv()} macro; and, b) the kernel variable argument list is terminated with a \texttt{NULL} sentinel.

The \progvar{krng} kernel, invoked in a loop to generate new batches of pseudo-random numbers, accepts three arguments, the first of which (number of PRNG states per loop iteration) remains constant. As such it only needs to be set once. The other two arguments are the buffers containing the PRNG states, used in a double-buffering fashion, as explained in the previous section. Thus, for each new invocation of the kernel, these arguments are swapped in order to achieve the double-buffering effect. The version using the OpenCL C host API (Listing~S1) sets the first (constant) argument in lines 402--403, and updates the second and third arguments within the loop in lines 418--424. Considering the first argument, cf4ocl provides no immediate advantages:

\begin{lstlisting}[firstnumber=264]
/* Set fixed argument of RNG kernel (number of random numbers in buffer). */
ccl_kernel_set_arg(krng, 0, ccl_arg_priv(bufs.numrn, cl_uint));
\end{lstlisting}

\noindent However, the second and third arguments are set in the same function call which also invokes the kernel:

\begin{lstlisting}[firstnumber=284]
/* Run random number generation kernel. */
evt_exec = ccl_kernel_set_args_and_enqueue_ndrange(krng, cq_main, 1,
	NULL, (const size_t*) &gws2, (const size_t*) &lws2, NULL, &err,
	ccl_arg_skip, bufdev1, bufdev2, /* Kernel arguments. */
	NULL);
\end{lstlisting}

\noindent Since it is not required to set the first argument again, we use the \progvar{ccl\_arg\_skip} constant to skip it.

The OpenCL host API allows for detailed profiling of OpenCL events. An event is generated each time an operation is enqueued in a command queue. These include, for example, kernel execution, memory transfers or image manipulations. If a command queue was instantiated with profiling capabilities, the associated events can be queried for profiling data. Thus, a profiling analysis requires the developer to keep all generated event objects, query them one-by-one, and perform the relevant calculations in order to obtain useful information. The section of code devoted to profiling in the pure OpenCL implementation of the example is shown in lines 455--523. As already discussed, this code does not determine event overlaps. The equivalent code for the cf4ocl implementation is shown in lines 309--335, and includes overlap calculation. Of these, the following should be highlighted:

\begin{lstlisting}[firstnumber=314, belowskip=4pt]
/* Add queues to the profiler object. */
ccl_prof_add_queue(prof, "Main", cq_main);
ccl_prof_add_queue(prof, "Comms", bufs.cq);
\end{lstlisting}
\begin{lstlisting}[firstnumber=318, belowskip=4pt, aboveskip=0pt]
/* Perform profiling calculations. */
ccl_prof_calc(prof, &err);
\end{lstlisting}
\begin{lstlisting}[firstnumber=322, aboveskip=0pt]
/* Show profiling info. */
fprintf(stderr, "%s", ccl_prof_get_summary(prof,
	CCL_PROF_AGG_SORT_TIME | CCL_PROF_SORT_DESC,
	CCL_PROF_OVERLAP_SORT_DURATION | CCL_PROF_SORT_DESC));
\end{lstlisting}

\begin{figure}[t]
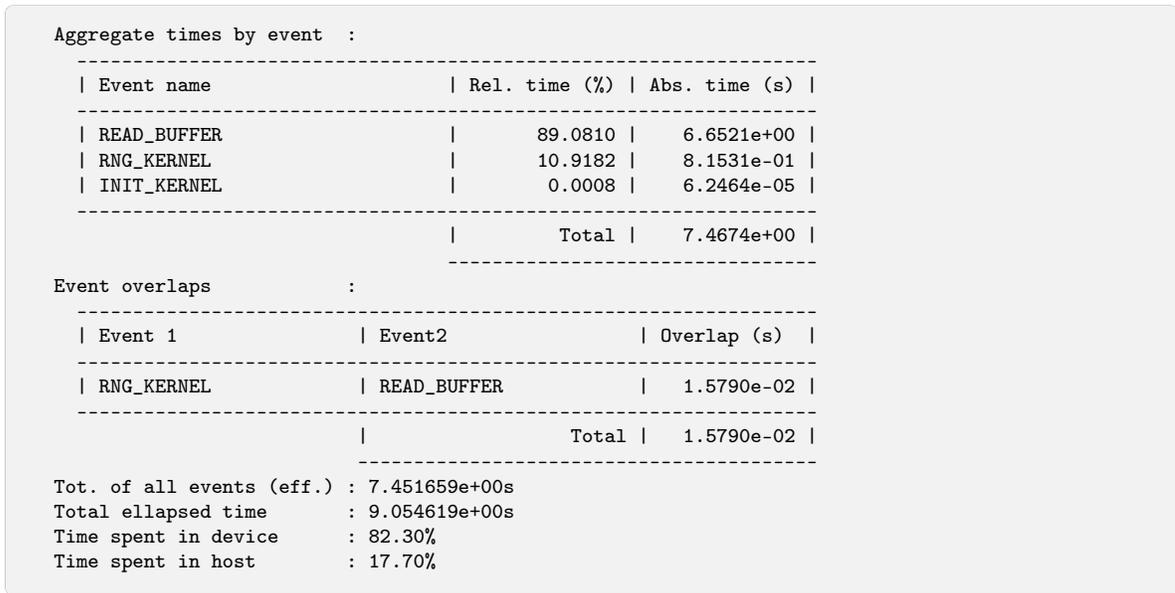

\begin{tcolorbox}[colbacklower=white, boxrule=0pt,colback=black!5,frame hidden]%
\begin{Verbatim}[fontsize=\footnotesize]
 Aggregate times by event  :
   ------------------------------------------------------------------
   | Event name                     | Rel. time (%) | Abs. time (s) |
   ------------------------------------------------------------------
   | READ_BUFFER                    |       89.0810 |    6.6521e+00 |
   | RNG_KERNEL                     |       10.9182 |    8.1531e-01 |
   | INIT_KERNEL                    |        0.0008 |    6.2464e-05 |
   ------------------------------------------------------------------
                                    |         Total |    7.4674e+00 |
                                    ---------------------------------
 Event overlaps            :
   ------------------------------------------------------------------
   | Event 1                | Event2                 | Overlap (s)  |
   ------------------------------------------------------------------
   | RNG_KERNEL             | READ_BUFFER            |   1.5790e-02 |
   ------------------------------------------------------------------
                            |                  Total |   1.5790e-02 |
                            -----------------------------------------
 Tot. of all events (eff.) : 7.451659e+00s
 Total ellapsed time       : 9.054619e+00s
 Time spent in device      : 82.30%
 Time spent in host        : 17.70%
\end{Verbatim}
\end{tcolorbox}
\caption{Possible profiling summary generated by the \progfunc{ccl\_prof\_get\_summary()} function for the cf4ocl implementation of the PRNG example.}
\label{fig:profsummary}
\end{figure}

\noindent Here, \progvar{prof} is an object of type \ooclass{CCLProf}, used for profiling OpenCL events with cf4ocl. In lines 315--316 the two command queues (type \ooclass{CCLQueue}) are added to this profiler object. With cf4ocl, the queues maintain a list of all event objects, thus it is not necessary for the developer to keep track of such objects. After all the utilized queues are added, the profiling calculations are performed with a call to \progfunc{ccl\_prof\_calc()} (line 319). As discussed in Section~\ref{sec:impl:prof}, different types of profiling information become available, and can be directly accessed or iterated over. However, a summary of the profiling analysis, as provided by the \progfunc{ccl\_prof\_get\_summary()} function, is sufficient for most purposes (lines 323--325). The flags indicate how to sort the aggregated events and overlaps, namely by time and duration, respectively. A possible output from this function is shown in Figure~\ref{fig:profsummary}.

\subsection{Comparison of performance} 

A crucial aspect of a wrapper library such as cf4ocl is the computational overhead it introduces. As such, we have tested both implementations with different conditions in order to determine possible overheads. In these tests, output was discarded by redirecting \textit{stdout} to the null device. Consequently, the \texttt{out} processes shown in Figure \ref{fig:exampleprng} have a very short duration. Furthermore, profiling is activated in both implementations. This constitutes a worst-case scenario for cf4ocl, since the additional overlap calculation step undertaken by its profiler is computationally expensive. The tests were performed under the following scenarios:

\begin{description}
\item[GPUs:] Nvidia GTX 1080 (Ubuntu 16.04, Intel Xeon E5-2650v3), AMD HD 7970 (Ubuntu 14.04, Intel Core i7-3930K) 
\item[Random numbers per iteration:] $n=2^{12},2^{14},2^{16},\ldots,2^{24}$
\item[Iterations:] $i=10^2,10^3,10^4$ 
\end{description}

\begin{figure}[t]
	\centering
	
	\subfloat[AMD HD7970.]{\includegraphics[width=1\linewidth]{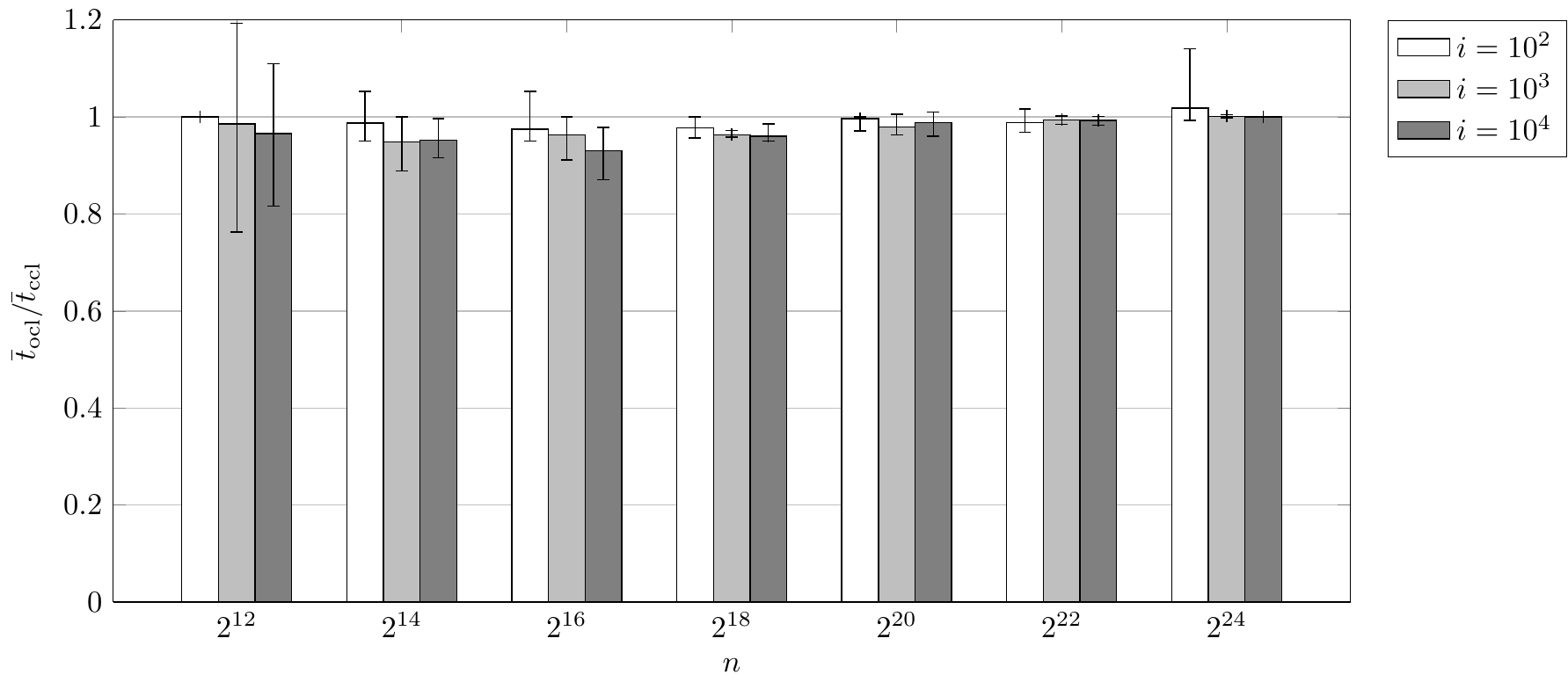}} \\
	\subfloat[Nvidia GTX 1080.]{\includegraphics[width=1\linewidth]{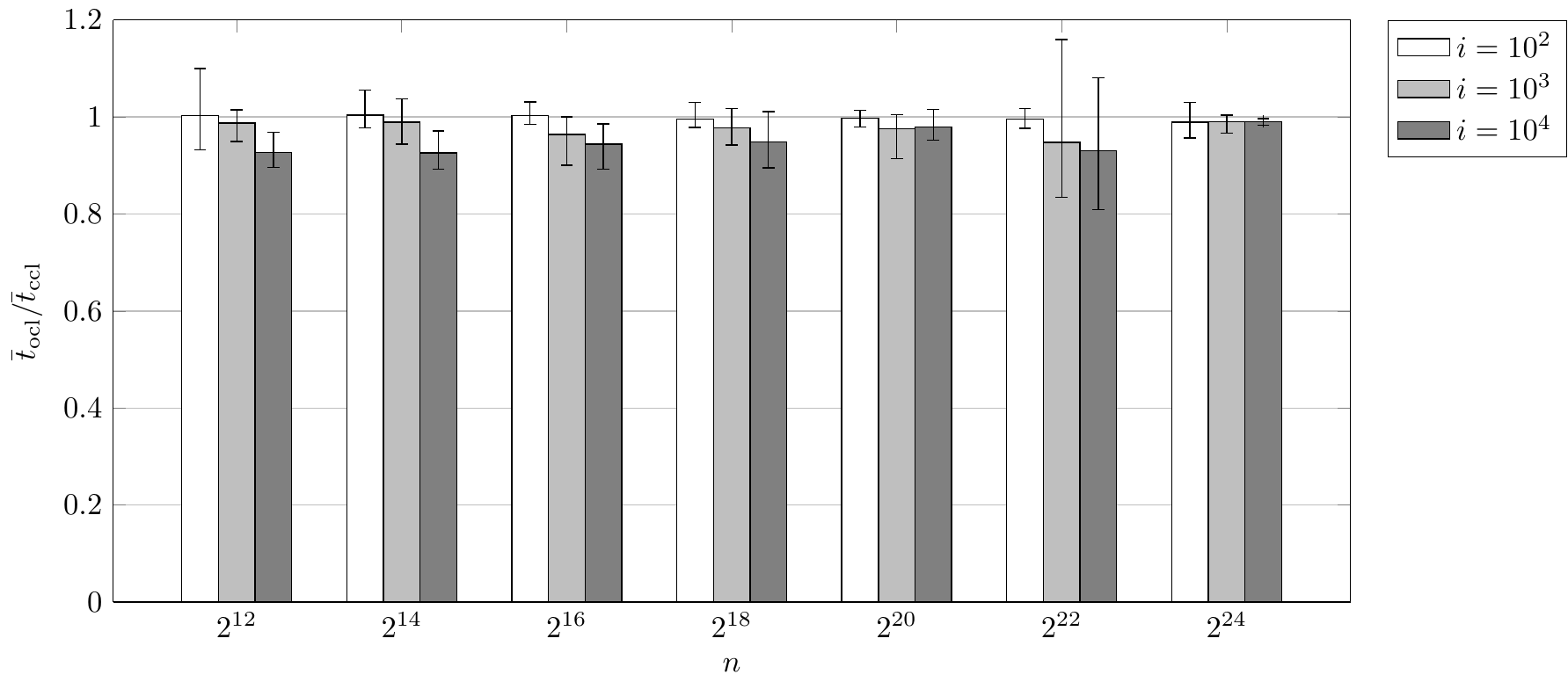}}
	\caption{Overhead of the cf4ocl example implementation against the pure OpenCL realization. Error bars show the relative maximum and minimum overheads. A total of 10 runs per parameter combination were performed for each implementation, with the maximum and minimum run times removed (thus, the results shown correspond to the remaining 8 runs). Overheads are determined by dividing $\overline{t}_{\text{ocl}}$, the average run time of the pure OpenCL realization, by $\overline{t}_{\text{ccl}}$, the average run time of the cf4ocl implementation. $n$ is the number of 64-bit random values generated per iteration, and $i$ is the total number of iterations. Figures were generated with PerfAndPubTools \cite{fachada2016perfandpubtools}.}
	\label{fig:empresults}
\end{figure}

Figure~\ref{fig:empresults} shows the average overhead of the cf4ocl implementation against the pure OpenCL realization. It is visible that cf4ocl does indeed introduce a small overhead, although in some cases manages to have similar or better average performance than the direct approach. Generally, the cf4ocl implementation performs better in relative terms when tested with fewer iterations, especially when $i=10^2$. This is to be expected, since fewer iterations imply less events to be analyzed, and consequently the overlap calculation step becomes less expensive. The graphs also seem to show that, for larger values of $n$, the overhead becomes negligible. This is also not surprising since a larger $n$ will necessarily involve more OpenCL computational work for the same number of iterations. Summarizing, a larger $n$ masks the profiling overhead, while a larger $i$ tends to expose it due to more events being generated.

\begin{figure}[t]
	\centering
	\includegraphics[width=1\linewidth]{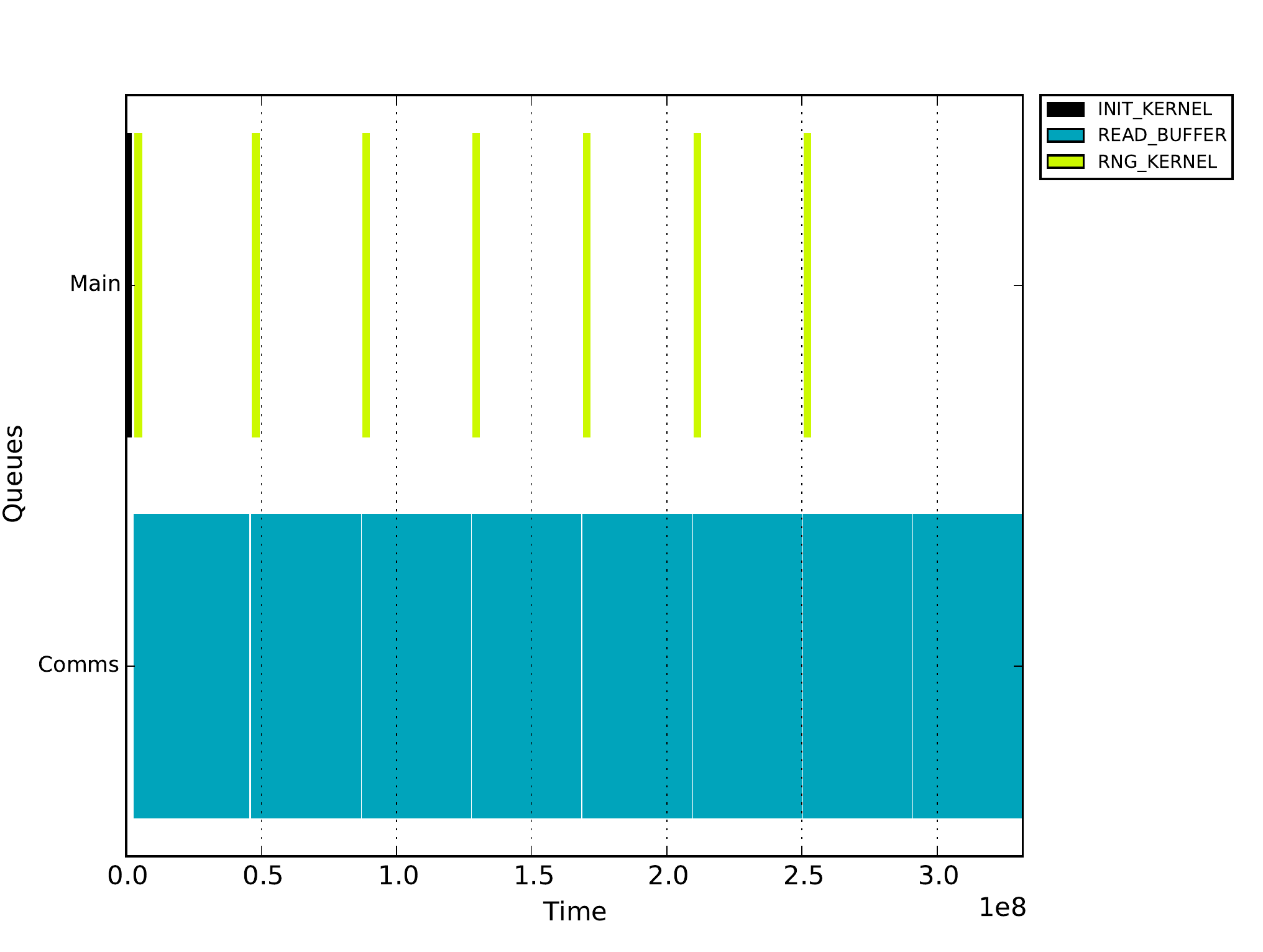}
	\caption{Queue utilization chart generated with the \texttt{ccl\_plot\_events} script for $n=2^{24}$ and $i=8$. Time is given in nanoseconds.}
	\label{fig:examplequeues}
\end{figure}

Cf4ocl provides a function to export profiling information to a file, which can then be analyzed with external tools. The \texttt{ccl\_plot\_events} script, bundled with cf4ocl, directly works with the exported file, plotting a queue utilization chart of the OpenCL commands executed in each queue. Figure \ref{fig:examplequeues} shows such a plot for the presented example when generating $2^{24}$ 64-bit random values per iteration during 8 iterations.

Since \textit{stdout} is redirected to the null device, the \texttt{out} process does not really show in the utilization chart. These kind of plots are very illustrative of application behavior. This plot in particular shows that the PRNG kernel invocations are indeed overlapping with device-host transfers, but are nonetheless much shorter. Thus, a more complex PRNG could probably be used instead, improving the quality of the generated random numbers, without compromising the program's efficiency.

\section{Conclusions}
\label{sec:conclusions}

In this paper we presented cf4ocl, a library with the aim of simplifying and accelerating the development of OpenCL host programs in C. Two implementations of a conceptual application example were presented and analyzed. The first was directly implemented with the OpenCL C host API, while the second, realized with cf4ocl, was shown to be more compact and straightforward. The overhead introduced by cf4ocl was small in a worst-case scenario, with a tendency to disappear in more compute heavy contexts. Furthermore, the cf4ocl profiling tools allowed for immediate inspection of command queue utilization, providing hints on how to optimize the example application.

\bibliographystyle{elsarticle-num}

\clearpage
\section*{Required Metadata}

\section*{Current executable software version}

\begin{table}[!h]
\begin{tabular}{|l|p{6.5cm}|p{6.5cm}|}
\hline
\textbf{Nr.} & \textbf{(executable) Software metadata description} & \textbf{Software metadata} \\
\hline
S1 & Current software version & 2.1.0 \\
\hline
S2 & Permanent link to executables of this version  & \url{https://github.com/fakenmc/cf4ocl/releases/tag/v2.1.0} \\
\hline
S3 & Legal Software License & LGPL-3.0 (library) and GPL-3.0 (utilities) \\
\hline
S4 & Computing platform/Operating System & Linux, macOS, Microsoft Windows, BSD, Unix-like \\
\hline
S5 & Installation requirements & GLib $\geqslant$ 2.32, OpenCL ICD $\geqslant$ 2.0 \\
\hline
S6 & User manual & \url{http://www.fakenmc.com/cf4ocl/docs/v2.1.0} \\
\hline
S7 & Support email for questions & \verb+nfachada@laseeb.org+ \\
\hline
\end{tabular}
\caption{Software metadata.}
\end{table}

\section*{Current code version}

\begin{table}[!h]
\begin{tabular}{|l|p{6.5cm}|p{6.5cm}|}
\hline
\textbf{Nr.} & \textbf{Code metadata description} & \textbf{Code metadata} \\
\hline
C1 & Current code version & v2.1.0 \\
\hline
C2 & Permanent link to code/repository used of this code version & \url{https://github.com/fakenmc/cf4ocl} \\
\hline
C3 & Legal Code License   & LGPL-3.0 (library) and GPL-3.0 (utilities, examples, tests, aux. scripts) \\
\hline
C4 & Code versioning system used & Git \\
\hline
C5 & Software code languages, tools, and services used & C, OpenCL, Bash, Python, CMake \\
\hline
C6 & Compilation requirements, operating environments & GLib $\geqslant$ 2.32, OpenCL ICD $\geqslant$ 1.0, CMake $\geqslant$ 2.8.3, C99 compiler, Git (optional), Bash and common GNU utilities (optional) \\
\hline
C7 & Developer documentation & \url{http://www.fakenmc.com/cf4ocl/docs/v2.1.0} \\
\hline
C8 & Support email for questions & \verb+nfachada@laseeb.org+ \\
\hline
\end{tabular}
\caption{Code metadata.}
\end{table}



\section*{Supplementary material}

## Description

This document contains Supplementary material for the manuscript "cf4ocl: a C framework for OpenCL" by Nuno Fachada, Vitor V. Lopes, Rui C. Martins and Agostinho C. Rosa. More specifically, this document contains the code listings for the example presented in the manuscript. The code is made available under the GNU General Public License, version 3.

## Code listings

**Listing S1** (`rng_ocl.c`) – Implementation of the PRNG example directly using the OpenCL API.

```c
#if defined(__APPLE__) || defined(__MACOSX)
  #include <OpenCL/opencl.h>
#else
  #include <CL/opencl.h>
#endif
#include <pthread.h>
#include <assert.h>
#include <sys/time.h>
#include <stdio.h>
#include <stdlib.h>
#include "cp_sem.h"

/* Define command queue flags depending on whether the profiling compile-time
 * flag set is set or not. */
#ifdef WITH_PROFILING
  #define CQ_FLAGS CL_QUEUE_PROFILING_ENABLE
#else
  #define CQ_FLAGS 0
#endif

/* Number of random number in buffer at each time.*/
#define NUMRN_DEFAULT 16777216

/* Number of iterations producing random numbers. */
#define NUMITER_DEFAULT 10000

/* Error handling macro. */
#define HANDLE_ERROR(status) \
  do { if (status != CL_SUCCESS) { \
    fprintf(stderr, "\nOpenCL error %d at line %d\n", status, __LINE__); \
    exit(EXIT_FAILURE); } \
  } while(0)

/* Kernels. */
#define KERNEL_INIT "init"
#define KERNEL_RNG "rng"
const char* kernel_filenames[] = { KERNEL_INIT ".cl", KERNEL_RNG ".cl" };

/* Thread semaphores. */
cp_sem_t sem_rng;
cp_sem_t sem_comm;

/* Information shared between main thread and data transfer/output thread. */
struct bufshare {

  /* Host buffer. */
  cl_ulong * bufhost;

  /* Device buffers. */
  cl_mem bufdev1;
  cl_mem bufdev2;

  /* Command queue for data transfers. */
  cl_command_queue cq;

  /* Array of RNG kernel and memory transfer events. */
  cl_event * evts;

  /* Possible transfer error. */
  cl_int status;
```

```
61
62     /* Number of random numbers in buffer. */
63     cl_uint numrn;
64
65     /* Number of iterations producing random numbers. */
66     unsigned int numiter;
67
68     /* Buffer size in bytes. */
69     size_t bufsize;
70
71  };
72
73  /* Write random numbers directly (as binary) to stdout. */
74  void * rng_out(void * arg) {
75
76     /* Increment aux variable. */
77     unsigned int i;
78
79     /* Buffer pointers. */
80     cl_mem bufdev1, bufdev2, bufswp;
81
82     /* Unwrap argument. */
83     struct bufshare * bufs = (struct bufshare *) arg;
84
85     /* Get initial buffers. */
86     bufdev1 = bufs->bufdev1;
87     bufdev2 = bufs->bufdev2;
88
89     /* Read random numbers and write them to stdout. */
90     for (i = 0; i < bufs->numiter; i++) {
91
92        /* Wait for RNG kernel from previous iteration before proceding with
93         * next read. */
94        cp_sem_wait(&sem_rng);
95
96        /* Read data from device buffer into host buffer. */
97        bufs->status = clEnqueueReadBuffer(bufs->cq, bufdev1, CL_TRUE, 0,
98           bufs->bufsize, bufs->bufhost, 0, NULL, &bufs->evts[i * 2]);
99
100       /* Signal that read for current iteration is over. */
101       cp_sem_post(&sem_comm);
102
103       /* If error occured in read, terminate thread and let main thread
104        * handle error. */
105       if (bufs->status != CL_SUCCESS) return NULL;
106
107       /* Write raw random numbers to stdout. */
108       fwrite(bufs->bufhost, sizeof(cl_ulong), (size_t) bufs->numrn, stdout);
109       fflush(stdout);
110
111       /* Swap buffers. */
112       bufswp = bufdev1;
113       bufdev1 = bufdev2;
114       bufdev2 = bufswp;
115
116    }
117
118    /* Bye. */
119    return NULL;
120 }
121
122 /**
123  * Main program.
124  *
125  * @param argc Number of command line arguments.
126  * @param argv Vector of command line arguments.
127  * @return `EXIT_SUCCESS` if program terminates successfully, or another
128  * `EXIT_FAILURE` if an error occurs.
129  * */
130 int main(int argc, char **argv) {
131
132    /* Aux. variable for loops. */
133    unsigned int i;
134
135    /* Host buffer. */
136    struct bufshare bufs = { NULL, NULL, NULL, NULL, NULL, 0, 0, 0, 0 };
137
```

```c
138    /* Communications thread. */
139    pthread_t comms_th;
140
141    /* OpenCL objects. */
142    cl_context ctx = NULL;
143    cl_device_id dev = NULL;
144    cl_program prg = NULL;
145    cl_kernel kinit = NULL, krng = NULL;
146    cl_command_queue cq_main = NULL;
147    cl_mem bufdev1 = NULL, bufdev2 = NULL, bufswp = NULL;
148    cl_event evt_kinit = NULL;
149    cl_platform_id * platfs = NULL;
150
151    /* Context properties. */
152    cl_context_properties ctx_prop[3] = { CL_CONTEXT_PLATFORM, 0, 0 };
153
154    /* Number of platforms. */
155    cl_uint nplatfs;
156
157    /* Number of devices in platform. */
158    cl_uint ndevs;
159
160    /* Device name. */
161    char* dev_name;
162
163    /* Status flag. */
164    cl_int status;
165
166    /* Size of information returned by clGet*Info functions. */
167    size_t infosize;
168
169    /* Generic vector where to put information returned by clGet*Info
170     * functions. */
171    void * info = NULL;
172
173    /* Real and kernel work sizes. */
174    size_t rws, gws1, gws2, lws1, lws2;
175
176    /* File pointer for files containing kernels. */
177    FILE * fp;
178
179    /* Length of kernel source code. */
180    size_t klens[2];
181
182    /* Kernel sources. */
183    char * ksources[2] = { NULL, NULL };
184
185    /* Variables for measuring execution time. */
186    struct timeval time1, time0;
187    double dt = 0;
188 #ifdef WITH_PROFILING
189    cl_ulong tstart, tend, tkinit = 0, tcomms = 0, tkrng = 0;
190 #endif
191
192    /* Initialize semaphores. */
193    cp_sem_init(&sem_rng, 1);
194    cp_sem_init(&sem_comm, 1);
195
196    /* Did user specify a number of random numbers? */
197    if (argc >= 2) {
198       /* Yes, use it. */
199       bufs.numrn = atoi(argv[1]);
200       bufs.bufsize = bufs.numrn * sizeof(cl_ulong);
201    } else {
202       /* No, use defaults. */
203       bufs.numrn = NUMRN_DEFAULT;
204       bufs.bufsize = NUMRN_DEFAULT * sizeof(cl_ulong);
205    }
206    rws = (size_t) bufs.numrn;
207
208    /* Did user specify a number of iterations producing random numbers? */
209    if (argc >= 3) {
210       /* Yes, use it. */
211       bufs.numiter = atoi(argv[2]);
212    } else {
213       /* No, use defaults. */
214       bufs.numiter = NUMITER_DEFAULT;
```

```c
215    }
216
217    /* Determine number of OpenCL platforms. */
218    status = clGetPlatformIDs(0, NULL, &nplatfs);
219    HANDLE_ERROR(status);
220
221    /* Allocate memory for existing platforms. */
222    platfs = (cl_platform_id*) malloc(sizeof(cl_platform_id) * nplatfs);
223
224    /* Get existing OpenCL platforms. */
225    status = clGetPlatformIDs(nplatfs, platfs, NULL);
226    HANDLE_ERROR(status);
227
228    /* Cycle through platforms until a GPU device is found. */
229    for (i = 0; i < nplatfs; i++) {
230
231       /* Determine number of GPU devices in current platform. */
232       status = clGetDeviceIDs(platfs[i], CL_DEVICE_TYPE_GPU, 0, NULL, &ndevs);
233       if (status == CL_DEVICE_NOT_FOUND) continue;
234       else HANDLE_ERROR(status);
235
236       /* Was any GPU device found in current platform? */
237       if (ndevs > 0) {
238
239          /* If so, get first device. */
240          status = clGetDeviceIDs(
241             platfs[i], CL_DEVICE_TYPE_GPU, 1, &dev, NULL);
242          HANDLE_ERROR(status);
243
244          /* Set the current platform as a context property. */
245          ctx_prop[1] = (cl_context_properties) platfs[i];
246
247          /* No need to cycle any more platforms. */
248          break;
249       }
250    }
251
252    /* If no GPU device was found, give up. */
253    assert(dev != NULL);
254
255    /* Get device name. */
256    status = clGetDeviceInfo(dev, CL_DEVICE_NAME, 0, NULL, &infosize);
257    HANDLE_ERROR(status);
258    dev_name = (char *) malloc(infosize);
259    status = clGetDeviceInfo(
260       dev, CL_DEVICE_NAME, infosize, (void *) dev_name, NULL);
261    HANDLE_ERROR(status);
262
263    /* Create context. */
264    ctx = clCreateContext(ctx_prop, 1, &dev, NULL, NULL, &status);
265    HANDLE_ERROR(status);
266
267    /* Create command queues. Here we use the "old" queue constructor, which is
268     * deprecated in OpenCL >= 2.0, and may throw a warning if compiled against
269     * such OpenCL versions. In cf4ocl the appropriate constructor is invoked
270     * depending on the underlying platform and the OpenCL version cf4ocl was
271     * built against. */
272    cq_main = clCreateCommandQueue(ctx, dev, CQ_FLAGS, &status);
273    HANDLE_ERROR(status);
274
275    bufs.cq = clCreateCommandQueue(ctx, dev, CQ_FLAGS, &status);
276    HANDLE_ERROR(status);
277
278    /* Read kernel sources into strings. */
279    for (i = 0; i < 2; i++) {
280
281       fp = fopen (kernel_filenames[i], "rb");
282       assert(fp != NULL);
283       fseek (fp, 0, SEEK_END);
284       klens[i] = (size_t) ftell(fp);
285       fseek(fp, 0, SEEK_SET);
286       ksources[i] = malloc(klens[i]);
287       fread(ksources[i], 1, klens[i], fp);
288       fclose (fp);
289    }
290
291    /* Create program. */
```

```c
292      prg = clCreateProgramWithSource(ctx, 2, (const char **) ksources,
293        (const size_t *) klens, &status);
294      HANDLE_ERROR(status);
295
296      /* Build program. */
297      status = clBuildProgram(
298        prg, 1, (const cl_device_id *) &dev, NULL, NULL, NULL);
299
300      /* Print build log in case of error. */
301      if (status == CL_BUILD_PROGRAM_FAILURE) {
302
303        /* Get size of build log. */
304        status = clGetProgramBuildInfo(
305          prg, dev, CL_PROGRAM_BUILD_LOG, 0, NULL, &infosize);
306        HANDLE_ERROR(status);
307
308        /* Allocate space for build log. */
309        info = malloc(infosize);
310
311        /* Get build log. */
312        status = clGetProgramBuildInfo(
313          prg, dev, CL_PROGRAM_BUILD_LOG, infosize, info, NULL);
314        HANDLE_ERROR(status);
315
316        /* Show build log. */
317        fprintf(stderr, "Error building program: \n%s", (char *) info);
318
319        /* Release build log. */
320        free(info);
321
322        /* Stop program. */
323        exit(EXIT_FAILURE);
324
325      } else {
326
327        HANDLE_ERROR(status);
328
329      }
330
331      /* Create init kernel. */
332      kinit = clCreateKernel(prg, KERNEL_INIT, &status);
333      HANDLE_ERROR(status);
334
335      /* Create rng kernel. */
336      krng = clCreateKernel(prg, KERNEL_RNG, &status);
337      HANDLE_ERROR(status);
338
339      /* Determine work sizes for each kernel. This is a minimum LOC approach
340       * which requires OpenCL >= 1.1. It does not account for the number of
341       * possibilities considered by the ccl_kernel_suggest_worksizes() cf4ocl
342       * function, namely multiple dimensions, OpenCL 1.0, kernel information
343       * unavailable, etc. */
344      status = clGetKernelWorkGroupInfo(
345        kinit, dev, CL_KERNEL_PREFERRED_WORK_GROUP_SIZE_MULTIPLE,
346        sizeof(size_t), &lws1, NULL);
347      HANDLE_ERROR(status);
348      gws1 = ((rws / lws1) + (((rws % lws1) > 0) ? 1 : 0)) * lws1;
349
350      status = clGetKernelWorkGroupInfo(
351        krng, dev, CL_KERNEL_PREFERRED_WORK_GROUP_SIZE_MULTIPLE,
352        sizeof(size_t), &lws2, NULL);
353      HANDLE_ERROR(status);
354      gws2 = ((rws / lws2) + (((rws % lws2) > 0) ? 1 : 0)) * lws2;
355
356      /* Allocate memory for host buffer. */
357      bufs.bufhost = (cl_ulong *) malloc(bufs.bufsize);
358
359      /* Create device buffers. */
360      bufdev1 = clCreateBuffer(
361        ctx, CL_MEM_READ_WRITE, bufs.bufsize, NULL, &status);
362      HANDLE_ERROR(status);
363
364      bufdev2 = clCreateBuffer(
365        ctx, CL_MEM_READ_WRITE, bufs.bufsize, NULL, &status);
366      HANDLE_ERROR(status);
367
368      /* Pass reference of device buffers to shared struct. */
```

```c
369      bufs.bufdev1 = bufdev1;
370      bufs.bufdev2 = bufdev2;
371
372      /* Initialize events buffers. */
373      bufs.evts = (cl_event *) calloc(2 * bufs.numiter - 1, sizeof(cl_event));
374
375      /* Print information. */
376      fprintf(stderr, "\n");
377      fprintf(stderr, " * Device name                    : %s\n", dev_name);
378      fprintf(stderr, " * Global/local work sizes (init): %u/%u\n",
379         (unsigned int) gws1, (unsigned int) lws1);
380      fprintf(stderr, " * Global/local work sizes (rng) : %u/%u\n",
381         (unsigned int) gws2, (unsigned int) lws2);
382      fprintf(stderr, " * Number of iterations           : %u\n",
383         (unsigned int) bufs.numiter);
384
385      /* Start profiling. */
386      gettimeofday(&time0, NULL);
387
388      /* Set arguments for initialization kernel. */
389      status = clSetKernelArg(
390         kinit, 0, sizeof(cl_mem), (const void *) &bufdev1);
391      HANDLE_ERROR(status);
392      status = clSetKernelArg(
393         kinit, 1, sizeof(cl_uint), (const void *) &bufs.numrn);
394      HANDLE_ERROR(status);
395
396      /* Invoke kernel for initializing random numbers. */
397      status = clEnqueueNDRangeKernel(cq_main, kinit, 1, NULL,
398         (const size_t *) &gws1, (const size_t *) &lws1, 0, NULL, &evt_kinit);
399      HANDLE_ERROR(status);
400
401      /* Set fixed argument of RNG kernel (number of random numbers in buffer). */
402      status = clSetKernelArg(
403         krng, 0, sizeof(cl_uint), (const void *) &bufs.numrn);
404      HANDLE_ERROR(status);
405
406      /* Wait for initialization to finish. */
407      status = clFinish(cq_main);
408      HANDLE_ERROR(status);
409
410      /* Invoke thread to output random numbers to stdout
411       * (in raw, binary form). */
412      pthread_create(&comms_th, NULL, rng_out, &bufs);
413
414      /* Produce random numbers. */
415      for (i = 0; i < bufs.numiter - 1; i++) {
416
417         /* Set RNG kernel arguments. */
418         status = clSetKernelArg(
419            krng, 1, sizeof(cl_mem), (const void *) &bufdev1);
420         HANDLE_ERROR(status);
421
422         status = clSetKernelArg(
423            krng, 2, sizeof(cl_mem), (const void *) &bufdev2);
424         HANDLE_ERROR(status);
425
426         /* Wait for read from previous iteration. */
427         cp_sem_wait(&sem_comm);
428
429         /* Handle possible errors in comms thread. */
430         HANDLE_ERROR(bufs.status);
431
432         /* Run random number generation kernel. */
433         status = clEnqueueNDRangeKernel(cq_main, krng, 1, NULL,
434            (const size_t *) &gws2, (const size_t *) &lws2, 0, NULL,
435            &bufs.evts[i * 2 + 1]);
436         HANDLE_ERROR(status);
437
438         /* Wait for random number generation kernel to finish. */
439         status = clFinish(cq_main);
440         HANDLE_ERROR(status);
441
442         /* Signal that RNG kernel from previous iteration is over. */
443         cp_sem_post(&sem_rng);
444
445         /* Swap buffers. */
```

```
      bufswp = bufdev1;
      bufdev1 = bufdev2;
      bufdev2 = bufswp;

    }

    /* Wait for output thread to finish. */
    pthread_join(comms_th, NULL);

    /* Stop profiling. */
    gettimeofday(&time1, NULL);

    /* Perform basic profiling calculations (i.e., we don't calculate overlaps,
     * which are automatically determined with the cf4ocl profiler). */

    /* Total time. */
    dt = time1.tv_sec - time0.tv_sec;
    if (time1.tv_usec >= time0.tv_usec)
      dt = dt + (time1.tv_usec - time0.tv_usec) * 1e-6;
    else
      dt = (dt-1) + (1e6 + time1.tv_usec - time0.tv_usec) * 1e-6;

    /* Show elapsed time. */
    fprintf(stderr, " * Total elapsed time                 : %es\n", dt);

#ifdef WITH_PROFILING

    /* Initialization kernel time. */
    status = clGetEventProfilingInfo(evt_kinit, CL_PROFILING_COMMAND_START,
      sizeof(cl_ulong), &tstart, NULL);
    HANDLE_ERROR(status);
    status = clGetEventProfilingInfo(evt_kinit, CL_PROFILING_COMMAND_END,
      sizeof(cl_ulong), &tend, NULL);
    tkinit = tend - tstart;

    /* Communication time. */
    for (i = 0; i < bufs.numiter; i++) {

      status = clGetEventProfilingInfo(bufs.evts[i * 2],
        CL_PROFILING_COMMAND_START, sizeof(cl_ulong), &tstart, NULL);
      HANDLE_ERROR(status);
      status = clGetEventProfilingInfo(bufs.evts[i * 2],
        CL_PROFILING_COMMAND_END, sizeof(cl_ulong), &tend, NULL);
      HANDLE_ERROR(status);
      tcomms += tend - tstart;

    }

    /* RNG kernel time. */
    for (i = 0; i < bufs.numiter - 1; i++) {

      /* RNG kernel time. */
      status = clGetEventProfilingInfo(bufs.evts[i * 2 + 1],
        CL_PROFILING_COMMAND_START, sizeof(cl_ulong), &tstart, NULL);
      HANDLE_ERROR(status);
      status = clGetEventProfilingInfo(bufs.evts[i * 2 + 1],
        CL_PROFILING_COMMAND_END, sizeof(cl_ulong), &tend, NULL);
      HANDLE_ERROR(status);
      tkrng += tend - tstart;

    }

    /* Show basic profiling info. */
    fprintf(stderr, " * Total time in 'init' kernel        : %es\n",
      (double) (tkinit * 1e-9));
    fprintf(stderr, " * Total time in 'rng' kernel         : %es\n",
      (double) (tkrng * 1e-9));
    fprintf(stderr, " * Total time fetching data from GPU : %es\n",
      (double) (tcomms * 1e-9));
    fprintf(stderr, "\n");

#endif

    /* Destroy OpenCL objects. */
    if (evt_kinit) clReleaseEvent(evt_kinit);
    for (i = 0; i < bufs.numiter * 2 - 1; i++) {
      if (bufs.evts[i]) clReleaseEvent(bufs.evts[i]);
```

```
523      }
524      if (bufdev1) clReleaseMemObject(bufdev1);
525      if (bufdev2) clReleaseMemObject(bufdev2);
526      if (cq_main) clReleaseCommandQueue(cq_main);
527      if (bufs.cq) clReleaseCommandQueue(bufs.cq);
528      if (kinit) clReleaseKernel(kinit);
529      if (krng) clReleaseKernel(krng);
530      if (prg) clReleaseProgram(prg);
531      if (ctx) clReleaseContext(ctx);
532
533      /* Free platforms buffer. */
534      if (platfs) free(platfs);
535
536      /* Free event buffers. */
537      if (bufs.evts) free(bufs.evts);
538
539      /* Free host resources */
540      if (bufs.bufhost) free(bufs.bufhost);
541
542      /* Free device name. */
543      if (dev_name) free(dev_name);
544
545      /* Destroy semaphores. */
546      cp_sem_destroy(&sem_comm);
547      cp_sem_destroy(&sem_rng);
548
549      /* Bye. */
550      return EXIT_SUCCESS;
551
552 }
```

**Listing S2** (`rng_ccl.c`) – Implementation of the PRNG example with cf4ocl.

```
 1 #include <cf4ocl2.h>
 2 #include <pthread.h>
 3 #include <assert.h>
 4 #include "cp_sem.h"
 5
 6 /* Define command queue flags depending on whether the profiling compile-time
 7  * flag set is set or not. */
 8 #ifdef WITH_PROFILING
 9     #define CQ_FLAGS CL_QUEUE_PROFILING_ENABLE
10 #else
11     #define CQ_FLAGS 0
12 #endif
13
14 /* Number of random number in buffer at each time.*/
15 #define NUMRN_DEFAULT 16777216
16
17 /* Number of iterations producing random numbers. */
18 #define NUMITER_DEFAULT 10000
19
20 /* Error handling macro. */
21 #define HANDLE_ERROR(err) \
22     do { if ((err) != NULL) { \
23         fprintf(stderr, "\nError at line %d: %s\n", __LINE__, (err)->message); \
24         ccl_err_clear(&(err)); \
25         exit(EXIT_FAILURE); } \
26     } while(0)
27
28 /* Kernels. */
29 #define KERNEL_INIT "init"
30 #define KERNEL_RNG "rng"
31 const char* kernel_filenames[] = { KERNEL_INIT ".cl", KERNEL_RNG ".cl" };
32
33 /* Thread semaphores. */
34 cp_sem_t sem_rng;
35 cp_sem_t sem_comm;
36
37 /* Information shared between main thread and data transfer/output thread. */
38 struct bufshare {
39
40     /* Host buffer. */
41     cl_ulong * bufhost;
42
```

```
 43    /* Device buffers. */
 44    CCLBuffer * bufdev1;
 45    CCLBuffer * bufdev2;
 46
 47    /* Command queue for data transfers. */
 48    CCLQueue * cq;
 49
 50    /* Possible transfer error. */
 51    CCLErr * err;
 52
 53    /* Number of random numbers in buffer. */
 54    cl_uint numrn;
 55
 56    /* Number of iterations producing random numbers. */
 57    unsigned int numiter;
 58
 59    /* Buffer size in bytes. */
 60    size_t bufsize;
 61
 62 };
 63
 64 /* Write random numbers directly (as binary) to stdout. */
 65 void * rng_out(void * arg) {
 66
 67    /* Increment aux variable. */
 68    unsigned int i;
 69
 70    /* Buffer pointers. */
 71    CCLBuffer * bufdev1, * bufdev2, * bufswp;
 72
 73    /* Unwrap argument. */
 74    struct bufshare * bufs = (struct bufshare *) arg;
 75
 76    /* Get initial buffers. */
 77    bufdev1 = bufs->bufdev1;
 78    bufdev2 = bufs->bufdev2;
 79
 80    /* Read random numbers and write them to stdout. */
 81    for (i = 0; i < bufs->numiter; i++) {
 82
 83       /* Wait for RNG kernel from previous iteration before proceding with
 84        * next read. */
 85       cp_sem_wait(&sem_rng);
 86
 87       /* Read data from device buffer into host buffer. */
 88       ccl_buffer_enqueue_read(bufdev1, bufs->cq, CL_TRUE, 0,
 89          bufs->bufsize, bufs->bufhost, NULL, &bufs->err);
 90
 91       /* Signal that read for current iteration is over. */
 92       cp_sem_post(&sem_comm);
 93
 94       /* If error occured in read, terminate thread and let main thread
 95        * handle error. */
 96       if (bufs->err) return NULL;
 97
 98       /* Write raw random numbers to stdout. */
 99       fwrite(bufs->bufhost, sizeof(cl_ulong), (size_t) bufs->numrn, stdout);
100       fflush(stdout);
101
102       /* Swap buffers. */
103       bufswp = bufdev1;
104       bufdev1 = bufdev2;
105       bufdev2 = bufswp;
106
107    }
108
109    /* Bye. */
110    return NULL;
111 }
112
113 /**
114  * Main program.
115  *
116  * @param argc Number of command line arguments.
117  * @param argv Vector of command line arguments.
118  * @return `EXIT_SUCCESS` if program terminates successfully, or another
119  * `EXIT_FAILURE` if an error occurs.
```

```
120   * */
121  int main(int argc, char **argv) {
122
123    /* Aux. variable for loops. */
124    unsigned int i;
125
126    /* Host buffer. */
127    struct bufshare bufs = { NULL, NULL, NULL, NULL, NULL, 0, 0, 0 };
128
129    /* Communications thread. */
130    pthread_t comms_th;
131
132    /* cf4ocl wrappers. */
133    CCLContext * ctx = NULL;
134    CCLDevice * dev = NULL;
135    CCLProgram * prg = NULL;
136    CCLKernel * kinit = NULL, *krng = NULL;
137    CCLQueue * cq_main = NULL;
138    CCLBuffer * bufdev1 = NULL, * bufdev2 = NULL, * bufswp = NULL;
139    CCLEvent * evt_exec = NULL;
140
141    /* Profiler object. */
142    CCLProf* prof = NULL;
143
144    /* Error management objects. */
145    CCLErr * err = NULL, * err_bld;
146
147    /* Device name. */
148    char* dev_name;
149
150    /* Real and kernel work sizes. */
151    size_t rws = 0, gws1 = 0, gws2 = 0, lws1 = 0, lws2 = 0;
152
153    /* Program build log. */
154    const char * bldlog;
155
156    /* Initialize semaphores. */
157    cp_sem_init(&sem_rng, 1);
158    cp_sem_init(&sem_comm, 1);
159
160    /* Did user specify a number of random numbers? */
161    if (argc >= 2) {
162      /* Yes, use it. */
163      bufs.numrn = atoi(argv[1]);
164      bufs.bufsize = bufs.numrn * sizeof(cl_ulong);
165    } else {
166      /* No, use defaults. */
167      bufs.numrn = NUMRN_DEFAULT;
168      bufs.bufsize = NUMRN_DEFAULT * sizeof(cl_ulong);
169    }
170    rws = (size_t) bufs.numrn;
171
172    /* Did user specify a number of iterations producing random numbers? */
173    if (argc >= 3) {
174      /* Yes, use it. */
175      bufs.numiter = atoi(argv[2]);
176    } else {
177      /* No, use defaults. */
178      bufs.numiter = NUMITER_DEFAULT;
179    }
180
181    /* Setup OpenCL context with GPU device. */
182    ctx = ccl_context_new_gpu(&err);
183    HANDLE_ERROR(err);
184
185    /* Get device. */
186    dev = ccl_context_get_device(ctx, 0, &err);
187    HANDLE_ERROR(err);
188
189    /* Get device name. */
190    dev_name = ccl_device_get_info_array(dev, CL_DEVICE_NAME, char*, &err);
191    HANDLE_ERROR(err);
192
193    /* Create command queues. */
194    cq_main = ccl_queue_new(ctx, dev, CQ_FLAGS, &err);
195    HANDLE_ERROR(err);
196    bufs.cq = ccl_queue_new(ctx, dev, CQ_FLAGS, &err);
```

```
197    HANDLE_ERROR(err);
198
199    /* Create program. */
200    prg = ccl_program_new_from_source_files(ctx, 2, kernel_filenames, &err);
201    HANDLE_ERROR(err);
202
203    /* Build program. */
204    ccl_program_build(prg, NULL, &err);
205
206    /* Print build log in case of error. */
207    if ((err) && (err->code == CL_BUILD_PROGRAM_FAILURE)) {
208       bldlog = ccl_program_get_build_log(prg, &err_bld);
209       HANDLE_ERROR(err_bld);
210       fprintf(stderr, "Error building program: \n%s", bldlog);
211       exit(EXIT_FAILURE);
212    }
213    HANDLE_ERROR(err);
214
215    /* Get kernels. */
216    kinit = ccl_program_get_kernel(prg, KERNEL_INIT, &err);
217    HANDLE_ERROR(err);
218    krng = ccl_program_get_kernel(prg, KERNEL_RNG, &err);
219    HANDLE_ERROR(err);
220
221    /* Determine preferred work sizes for each kernel. */
222    ccl_kernel_suggest_worksizes(kinit, dev, 1, &rws, &gws1, &lws1, &err);
223    HANDLE_ERROR(err);
224    ccl_kernel_suggest_worksizes(krng, dev, 1, &rws, &gws2, &lws2, &err);
225    HANDLE_ERROR(err);
226
227    /* Allocate memory for host buffer. */
228    bufs.bufhost = (cl_ulong*) malloc(bufs.bufsize);
229
230    /* Create device buffers. */
231    bufdev1 = ccl_buffer_new(
232       ctx, CL_MEM_READ_WRITE, bufs.bufsize, NULL, &err);
233    HANDLE_ERROR(err);
234    bufdev2 = ccl_buffer_new(
235       ctx, CL_MEM_READ_WRITE, bufs.bufsize, NULL, &err);
236    HANDLE_ERROR(err);
237
238    /* Pass reference of device buffers to shared struct. */
239    bufs.bufdev1 = bufdev1;
240    bufs.bufdev2 = bufdev2;
241
242    /* Print information. */
243    fprintf(stderr, "\n");
244    fprintf(stderr, " * Device name                     : %s\n", dev_name);
245    fprintf(stderr, " * Global/local work sizes (init): %u/%u\n",
246       (unsigned int) gws1, (unsigned int) lws1);
247    fprintf(stderr, " * Global/local work sizes (rng) : %u/%u\n",
248       (unsigned int) gws2, (unsigned int) lws2);
249    fprintf(stderr, " * Number of iterations            : %u\n",
250       (unsigned int) bufs.numiter);
251
252    /* Start profiling. */
253    prof = ccl_prof_new();
254    ccl_prof_start(prof);
255
256    /* Invoke kernel for initializing random numbers. */
257    evt_exec = ccl_kernel_set_args_and_enqueue_ndrange(kinit, cq_main, 1, NULL,
258       (const size_t*) &gws1, (const size_t*) &lws1, NULL, &err,
259       bufdev1, ccl_arg_priv(bufs.numrn, cl_uint), /* Kernel arguments. */
260       NULL);
261    HANDLE_ERROR(err);
262    ccl_event_set_name(evt_exec, "INIT_KERNEL");
263
264    /* Set fixed argument of RNG kernel (number of random numbers in buffer). */
265    ccl_kernel_set_arg(krng, 0, ccl_arg_priv(bufs.numrn, cl_uint));
266
267    /* Wait for initialization to finish. */
268    ccl_queue_finish(cq_main, &err);
269    HANDLE_ERROR(err);
270
271    /* Invoke thread to output random numbers to stdout
272     * (in raw, binary form). */
273    pthread_create(&comms_th, NULL, rng_out, &bufs);
```

```c
274
275    /* Produce random numbers. */
276    for (i = 0; i < bufs.numiter - 1; i++) {
277
278       /* Wait for read from previous iteration. */
279       cp_sem_wait(&sem_comm);
280
281       /* Handle possible errors in comms thread. */
282       HANDLE_ERROR(bufs.err);
283
284       /* Run random number generation kernel. */
285       evt_exec = ccl_kernel_set_args_and_enqueue_ndrange(krng, cq_main, 1,
286          NULL, (const size_t*) &gws2, (const size_t*) &lws2, NULL, &err,
287          ccl_arg_skip, bufdev1, bufdev2, /* Kernel arguments. */
288          NULL);
289       HANDLE_ERROR(err);
290       ccl_event_set_name(evt_exec, "RNG_KERNEL");
291
292       /* Wait for random number generation kernel to finish. */
293       ccl_queue_finish(cq_main, &err);
294       HANDLE_ERROR(err);
295
296       /* Signal that RNG kernel from previous iteration is over. */
297       cp_sem_post(&sem_rng);
298
299       /* Swap buffers. */
300       bufswp = bufdev1;
301       bufdev1 = bufdev2;
302       bufdev2 = bufswp;
303
304    }
305
306    /* Wait for output thread to finish. */
307    pthread_join(comms_th, NULL);
308
309    /* Stop profiling. */
310    ccl_prof_stop(prof);
311
312 #ifdef WITH_PROFILING
313
314    /* Add queues to the profiler object. */
315    ccl_prof_add_queue(prof, "Main", cq_main);
316    ccl_prof_add_queue(prof, "Comms", bufs.cq);
317
318    /* Perform profiling calculations. */
319    ccl_prof_calc(prof, &err);
320    HANDLE_ERROR(err);
321
322    /* Show profiling info. */
323    fprintf(stderr, "%s", ccl_prof_get_summary(prof,
324       CCL_PROF_AGG_SORT_TIME | CCL_PROF_SORT_DESC,
325       CCL_PROF_OVERLAP_SORT_DURATION | CCL_PROF_SORT_DESC));
326 #else
327
328    /* Show elapsed time. */
329    fprintf(stderr, " * Total elapsed time                    : %es\n",
330       ccl_prof_time_elapsed(prof));
331
332 #endif
333
334    /* Destroy profiler object. */
335    ccl_prof_destroy(prof);
336
337    /* Destroy cf4ocl wrappers - only the ones created with ccl_*_new()
338     * functions. */
339    if (bufdev1) ccl_buffer_destroy(bufdev1);
340    if (bufdev2) ccl_buffer_destroy(bufdev2);
341    if (cq_main) ccl_queue_destroy(cq_main);
342    if (bufs.cq) ccl_queue_destroy(bufs.cq);
343    if (prg) ccl_program_destroy(prg);
344    if (ctx) ccl_context_destroy(ctx);
345
346    /* Free host resources */
347    if (bufs.bufhost) free(bufs.bufhost);
348
349    /* Destroy semaphores. */
350    cp_sem_destroy(&sem_comm);
```

```
351    cp_sem_destroy(&sem_rng);
352
353    /* Check that all cf4ocl wrapper objects are destroyed. */
354    assert(ccl_wrapper_memcheck());
355
356    /* Bye. */
357    return EXIT_SUCCESS;
358
359 }
```

**Listing S3** (`cp_sem.h`) – Compatibility header for cross-platform semaphore usage.

```
 1 #ifdef __APPLE__
 2 #include <dispatch/dispatch.h>
 3 #else
 4 #include <semaphore.h>
 5 #endif
 6
 7 /**
 8  * The semaphore object.
 9  * */
10 typedef struct {
11 #ifdef __APPLE__
12   dispatch_semaphore_t sem;
13 #else
14   sem_t sem;
15 #endif
16 } cp_sem_t;
17
18 /**
19  * Initialize semaphore.
20  * */
21 static inline void cp_sem_init(cp_sem_t * s, unsigned int val) {
22 #ifdef __APPLE__
23   unsigned int i;
24   s->sem = dispatch_semaphore_create(0);
25   for (i = 0; i < val; i++)
26     dispatch_semaphore_signal(s->sem);
27 #else
28   sem_init(&s->sem, 0, val);
29 #endif
30 }
31
32 /**
33  * Destroy semaphore.
34  * */
35 static inline void cp_sem_destroy(cp_sem_t * s) {
36 #ifdef __APPLE__
37   dispatch_release((dispatch_object_t) s->sem);
38 #else
39   sem_destroy(&s->sem);
40 #endif
41 }
42
43 /**
44  * Wait on semaphore if value is zero, otherwise decrement semaphore.
45  * */
46 static inline int cp_sem_wait(cp_sem_t * s) {
47   int res;
48 #ifdef __APPLE__
49   res = (int) dispatch_semaphore_wait(s->sem, DISPATCH_TIME_FOREVER);
50 #else
51   res = sem_wait(&s->sem);
52 #endif
53   return res;
54 }
55
56 /**
57  * Unlock semaphore.
58  * */
59 static inline void cp_sem_post(cp_sem_t * s) {
60 #ifdef __APPLE__
61   dispatch_semaphore_signal(s->sem);
62 #else
63   sem_post(&s->sem);
```

```
64  #endif
65  }
```

## Listing S4 (`init.cl`) – Initialization kernel.

```
 1  __kernel void init(
 2      __global uint2 *seeds,
 3      const uint nseeds) {
 4
 5    /* Global ID of current work-item. */
 6    size_t gid = get_global_id(0);
 7
 8    /* Does this work-item has anything to do? */
 9    if (gid < nseeds) {
10
11      /* Final value. */
12      uint2 final;
13
14      /* Initialize low bits with global ID. */
15      uint a = (uint) gid;
16
17      /* Scramble seed using hashes described in
18       * http://www.burtleburtle.net/bob/hash/integer.html */
19
20      /* Low bits. */
21      a = (a + 0x7ed55d16) + (a << 12);
22      a = (a ^ 0xc761c23c) ^ (a >> 19);
23      a = (a + 0x165667b1) + (a << 5);
24      a = (a + 0xd3a2646c) ^ (a << 9);
25      a = (a + 0xfd7046c5) + (a << 3);
26      a = (a ^ 0xb55a4f09) ^ (a >> 16);
27
28      /* Keep low bits. */
29      final.x = a;
30
31      /* High bits. */
32      a = (a ^ 61) ^ (a >> 16);
33      a = a + (a << 3);
34      a = a ^ (a >> 4);
35      a = a * 0x27d4eb2d;
36      a = a ^ (a >> 15);
37
38      /* Keep high bits. */
39      final.y = a;
40
41      /* Save random number in buffer. */
42      seeds[gid] = final;
43
44    }
45  }
```

## Listing S5 (`rng.cl`) – Pseudo-random number generation kernel.

```
 1  __kernel void rng(
 2      const uint nseeds,
 3      __global ulong *in,
 4      __global ulong *out) {
 5
 6    /* Global ID of current work-item. */
 7    size_t gid = get_global_id(0);
 8
 9    /* Does this work-item has anything to do? */
10    if (gid < nseeds) {
11
12      /* Fetch current state. */
13      ulong state = in[gid];
14
15      /* Update state using simple xor-shift RNG. */
16      state ^= (state << 21);
17      state ^= (state >> 35);
18      state ^= (state << 4);
19
20      /* Save new state in buffer. */
```

```
21      out[gid] = state;
22
23   }
24 }
```



\end{document}